\documentclass[prd,aps,nofootinbib,floatfix,10 pt]{revtex4}
\usepackage{amssymb}
\usepackage{amsmath,graphicx,color,epsfig}

\begin{document}

\title{Electroweak unification of quarks and leptons in a gauge group $SU_{C}(3)\times SU(4)\times U_{X}(1)$}
\author{Fayyazuddin\footnote{fayyazuddins@gmail.com}}
\affiliation{National Centre for Physics and Physics Department,\\
Quaid-i-Azam University, Islamabad 45320, Pakistan.}
\date{\today}

\begin{abstract}
A model for electroweak unification of quarks and leptons, in a
gauge group $SU_{C}(3)\times SU(4)\times U_{X}(1)$ is constructed.
The model requires, three generations of quarks and leptons which
are replicas (mirror) of the standard quarks and leptons. The gauge
group $SU(4)\times U_{X}(1)$ is broken in such a way so as to
reproduce standard model and to generate heavy masses for the vector
bosons ($W^{\pm}_{R_{\mu}}$,
$Z^{\prime}_{\mu}$,$Z^{\prime\prime}_{\mu}$), the leptoquarks and
mirror fermions. It is shown lower limit on mass scale of mirror
fermions is $m_{E}\geq\frac{m_{Z}}{2}$, $E^{-}$ being the lightest
mirror fermion coupled to Z boson. As the universe expands, the
heavy matter is decoupled at an early stage of expansion and may be
a source of dark matter. Leptoquarks in the model connect the
standard model and mirror fermions. Baryon genesis
in our Universe implies antibaryon genesis in mirror Universe.
\end{abstract}
 \maketitle

\section{Introduction}
A unification model based on the gauge group SU(4), with
$\sin^2\theta_{W}=1/4$ at the unification mass scale 5.8TeV, was
proposed in 1984 \cite{1}. Although leptons of standard model can be
assigned to the representation $4_R$ or equivalently $4^{\ast}_{L}$,
the cancelation of anomaly requires Hahn-Nambu quarks \cite{2} with
integral charges. Several models based on the extension of SU(4) to
the semi-simple group $SU_C(3)\times SU(4)\times U(1)$ have been
constructed \cite{3}-\cite{7}. However all these models require
extra quarks and leptons beyond those of standard model group
$SU_C(3)\times SU_L(2)\times U_Y(1)$.

By extending the group $SU_L(2)\times U_Y(1)$ to $SU(4)\times
U_{X}(1)$, a model is proposed in this paper to unify two separate
multiplets of quarks and leptons of standard model in a single
multiplet of electoweak unification group $SU(4)\times U_X(1)$.
However the unification requires pairing of three generations of
quarks (leptons) of the standard model with three generations of
mirror leptons (quarks) in the fundamental representation
$4_L$ or $4_R$ of SU(4).

It is an attempt for electroweak  unification of quarks and leptons
in contrast to unification of quarks and leptons of standard model
in a grand unification group G $\subset SU_C(3)\times SU_L(2)\times
U_Y(1)$.

The grand unification group SU(5) is most economical; quarks and
leptons of standard model are assigned to the representations
$\overline{5}+10$ of SU(5). Out of twenty four vector bosons of SU(5),
eight are gluons, four bosons of
standard model and twelve are lepto-quarks. In SU(5) the selection
rule $\Delta (B-L)=0$ holds. The proton life time
$\tau_p(p\rightarrow e^+\pi^0)\sim4\times10^{29}$, predicted in
SU(5), contradicts the experimental limit $\tau_p(p\rightarrow
e^+\pi^0)\sim5\times10^{32}$. The extension of SU(5) to minimal
supersymmetric model (MSSM) SU(5),(to avoid the above problem and
for unification of three coupling constants of standard model at GUT
scale) is rich in proliferation of fermions and bosons with large
number of unknown parameters \cite{8}. It is hard to go beyond
standard model without proliferations of fermions and bosons. In
this respect, electroweak unification model formulated in this paper
is not different as it requires mirror fermions. Out of sixteen
vector bosons in this model, eight are lepto-quarks, eight are vector bosons ($W_{L\mu}^{\pm},Z_{\mu},A_{\mu}$) and ($W_{R\mu}^{\pm},Z_{\mu}^{\prime},Z^{\prime\prime}_{\mu}$). The
lepto-quarks connect the ordinary and mirror fermions.

Lower limit to the mass scale of mirror fermions is obtained from
the fact that decay width of Z-boson is accurately accounted for in
terms of the decay channels available in the standard model. Hence
the decay channel $Z\rightarrow E^-E^+$, involving the lightest
mirror fermions, gives the limit $m_E\geq m_Z/2$; assuming similar
mass hierarchy for mirror fermions as for standard model
fermions, one would expect left-handed mirror neutrinos $\nu_{Ei}$
to be stable and good candidates for dark matter. Another
feature of electroweak unification model is that matter dominated
our Universe implies antimatter dominated mirror Universe.
\section{Electroweak unification group $SU(4)\times U_x(1)$}

The group $SU(4)$ has $SU_{L}(2)\times SU_{R}(2)\times U_{Y_{1}}(1)$ as
its subgroup with three diagonal matrices:
\begin{eqnarray}
\tau _{3} &=&\lambda _{3}:\text{ diag}(1,-1,0,0)\notag \\
s_{3} &=&\frac{1}{\sqrt{3}}\left( -\lambda _{8}+\sqrt{2}\lambda
_{15}\right)
:\text{diag}(0,0,1,-1)\notag \\
Y_{1} &=&\frac{2}{3\sqrt{3}}\left( 2\lambda _{8}+\sqrt{2}\lambda
_{15}\right) \text{ :\thinspace\ diag}(2/3,2/3,-2/3,-2/3)\notag \\
Q(SU(4)) &=&\frac{1}{2}\left( \tau _{3}+s_{3} +Y_{1}\right) :\text{
\ \ diag}(5/6,-1/6,1/6,-5/6)
\end{eqnarray}
A charge structure of fermions not realized in nature. First we note that at unification scale, $sin^{2}\theta_{W}$ = $e^{2}/g^{2}$=$9/26$
$\approx$ $0.346$ to be compared with the $SU(5)$ value
$sin^{2}\theta_{W}$=$3/8$ $\approx$ $0.375$. To see this, we note
that
\begin{eqnarray}
c_{1}^{2}=\frac{1}{2}TrY_{1}^{2}=8/9
\end{eqnarray}
Let $B_{1\mu}$ be the vector boson associated with $U_{Y_{1}}(1)$
with the gauge coupling constant $g_{1}$:
\begin{eqnarray}
g_{1}^{2}=\frac{g^{2}}{c_{1}^{2}}=\frac{9g^{2}}{8}
\end{eqnarray}
Now
\begin{eqnarray}
\frac{1}{e^{2}}=\frac{2}{g^{2}}+\frac{1}{g_{1}^{2}}=\frac{26}{9}\frac{1}{g^{2}}
\end{eqnarray}
giving $sin^{2}\theta_{W}$ =$9/26$ at the unification mass scale.

With odd charge structure of fermions, the proposed $SU(4)$ is
not realized in nature; it must be extended to $SU(4)\times
U_{X}(1)$ $\supset SU_{L}(2)\times SU_{R}(2)\times U_{Y_{1}}\times
U_{X}(1)$

with
\begin{eqnarray}
Q=\frac{1}{2}(\tau_{3}+s_{3}+Y_{1}+Y_{X})
\end{eqnarray}
For
\begin{eqnarray}
Y_{X}=-1/3,Q=diag(2/3,-1/3,0,-1)\notag\\
Y_{X}=1/3,Q=diag(1,0,1/3,-2/3)
\end{eqnarray}
An attractive charge structure, suitable to assign quarks-leptons
and their CP-conjugates to the multiplet $\textbf{$4_{L}$}$ and
$\textbf{$4_{R}$}$ respectively. The three generations of ordinary
matter
\begin{eqnarray}
\psi _{q}=\left(
\begin{array}{c}
u_{i}^{a} \\
d_{i}^{a}%
\end{array}%
\right) ,\text{ }\psi _{l}=\left(
\begin{array}{c}
\nu _{ei} \\
e_{i}%
\end{array}%
\right)
\end{eqnarray}

are paired with their replicas
\begin{eqnarray}
\psi _{Q}=\left(
\begin{array}{c}
U_{i}^{a} \\
D_{i}^{a}%
\end{array}%
\right) ,\text{ }\psi _{E}=\left(
\begin{array}{c}
\nu _{Ei} \\
E_{i}%
\end{array}%
\right)
\end{eqnarray}
a: color index a =r,y,b\\
i: generation index i=1,2,3\\
The three generations of quarks and leptons are assigned to the
gauge group $SU(4)\times U_{X}(1)$ in the following two ways
(suppressing both color and generation indices $a$ and $i$.

Version 1:
\begin{eqnarray}
\text{\ }F_{L}=\left(
\begin{array}{c}
\psi _{q} \\
\psi _{E}%
\end{array}%
\right) _{L}=\left(
\begin{array}{c}
u \\
d \\
\nu _{E} \\
E%
\end{array}%
\right) _{L,Y_{X}=-1/3}f_{R}:[%
\begin{array}{c}
u_{R},d_{R}:Y_{X}=4/3,-2/3 \\
E_{R}:\text{ \ }Y_{X}=-2%
\end{array}%
\end{eqnarray}
\begin{eqnarray*}
\text{ \ \ \ \ }F_{R}=\left(
\begin{array}{c}
\psi _{l}^{c} \\
\psi _{Q}^{c}%
\end{array}%
\right) _{R}=\left(
\begin{array}{c}
e^{c} \\
-\nu _{e}^{c} \\
-D^{c} \\
U^{c}%
\end{array}%
\right) _{R,Y_{X}=1/3}f_{L}:[%
\begin{array}{c}
U_{L}^{c},D_{L}^{c}:Y_{X}^{c}=-4/3,2/3 \\
e_{L}^{c}:\text{ \ }Y_{X}^{c}=2%
\end{array}%
\end{eqnarray*}
Version 2:
\begin{eqnarray}
\text{ \ \ \ \ \ \ \ \ \ }F_{L}=\left(
\begin{array}{c}
u \\
d \\
\nu _{E} \\
E^{-}%
\end{array}%
\right) _{L,Y_{X}=-1/3},\text{ \ \ \ \ \ }F_{R}^{\prime }=\left(
\begin{array}{c}
U \\
D \\
N_{e} \\
e^{-}%
\end{array}%
\right) _{R,Y_{X}=-1/3}
\end{eqnarray}
\begin{eqnarray*}
\text{ \ \ \ \ \ }F_{R}=\left(
\begin{array}{c}
e^{+} \\
-\nu _{e}^{c} \\
-D^{c} \\
U^{c}%
\end{array}%
\right) _{R,Y_{X}=1/3},\text{ \ \ \ \ \ }F_{L}^{\prime }=\left(
\begin{array}{c}
E^{+} \\
-N_{E}^{c} \\
-d^{c} \\
u^{c}%
\end{array}%
\right) _{L,Y_{X}=1/3}
\end{eqnarray*}
 Both these representations are anomaly free as there are equal
numbers of left-handed and right handed multiplets. The second
version has complicated phenomenology, we will not consider it any
further.

The gauge symmetry is broken to $U_{em}(1)$ in such a way so as to
generate the mass patteren:
$m(\text{lepto-quarks})>>m^{\pm}_{W_R},m_{Z^{\prime\prime}},m_{Z^{\prime}}>>m_{W_L}^{\pm},m_Z$.
As the Universe expands, lepto-quarks are decoupled at an early
stage of expansion followed by vector bosons
$W_R^{\pm},Z^{\prime},Z^{\prime\prime}$ and mirror fermions: The
lepto-quarks bosons carry $\Delta B$ = $\pm 1/3$, $\Delta L$ = $\mp$
1, so that when color is taken into account, $\Delta(B+L)$ = 0, in
contrast to $SU(5)$ where the selection rule is $\Delta(B-L)$ = 0.
From Eqs. (9) and (10), it is clear that lepto-quarks bosons connect
the ordinary and mirror fermions.
\section{Gauge vector bosons and spontaneous symmetry breaking}
The gauge vector bosons of $SU(4)$ belong to adjoint representation.
The 16 vector bosons of $SU(4)\times U_{X}(1)$ can be written as
$4\times 4$ matrix.
\begin{eqnarray}
W_{\mu}\equiv \left(
\begin{array}{cccc}
W_{11}(W_{11}^{\prime })_{\mu } & \sqrt{2}W_{L\mu }^{-} &
\sqrt{2}X_{1\mu
}^{-2/3} & \sqrt{2}Y_{1\mu }^{-5/3} \\
\sqrt{2}W_{L\mu }^{+} & W_{22}(W_{22}^{\prime })_{\mu } &
\sqrt{2}X_{2\mu
}^{1/3} & \sqrt{2}Y_{2\mu }^{-2/3} \\
\sqrt{2}X_{1\mu }^{2/3} & \sqrt{2}X_{2\mu }^{-1/3} &
W_{33}(W_{33}^{\prime
})_{\mu } & \sqrt{2}W_{R\mu }^{-} \\
\sqrt{2}Y_{1\mu }^{5/3} & \sqrt{2}Y_{2\mu }^{2/3} & \sqrt{2}W_{R\mu
}^{+} &
W_{44}(W_{44}^{\prime })_{\mu }%
\end{array}%
\right)
\end{eqnarray}%
where
\begin{eqnarray}
W_{11}(W_{11}^{\prime })_{\mu } &=&W_{3L\mu }+\frac{2}{3}\frac{g_{1}}{g}%
B_{1\mu }\mp \frac{1}{3}\frac{g_{X}}{g}X_{\mu }\notag \\
W_{22}(W_{22}^{\prime })_{\mu } &=&-W_{3L\mu }+\frac{2}{3}\frac{g_{1}}{g}%
B_{1\mu }\mp \frac{1}{3}\frac{g_{X}}{g}X_{\mu }\notag \\
W_{33}(W_{33}^{\prime })_{\mu } &=&W_{3R\mu }-\frac{2}{3}\frac{g_{1}}{g}%
B_{1\mu }\mp \frac{1}{3}\frac{g_{X}}{g}X_{\mu }\notag \\
W_{44}(W_{44}^{\prime })_{\mu } &=&-W_{3R\mu }-\frac{2}{3}\frac{g_{1}}{g}%
B_{1\mu }\mp \frac{1}{3}\frac{g_{X}}{g}X_{\mu }
\end{eqnarray}
In addition, another combination of neutral vector bosons is
introduced
\begin{eqnarray}
W_{44\mu }^{\prime \prime}=-W_{3R\mu }-\frac{2}{3}\frac{g_{1}}{g}B_{1\mu }+%
\frac{5}{3}\frac{g_{X}}{g}X_{\mu }
\end{eqnarray}
$W_{44\mu }^{\prime \prime}$ is not coupled to fermions as is clear
from Eq.(12)

Now
\begin{eqnarray}
\frac{A_{\mu }}{e} &=&\frac{W_{3L\mu }}{g}+\frac{W_{3R\mu }}{g}+\frac{%
B_{1\mu }}{g_{1}}+\frac{X_{\mu }}{g_{X}}\notag \\
\frac{B_{\mu }}{g^{\prime }} &=&\frac{1}{g}W_{3R\mu
}+\frac{1}{g^{\prime \prime }}\left( \frac{g^{\prime \prime
}}{g_{1}}B_{1\mu }+\frac{g^{\prime \prime }}{g_{X}}X_{\mu
}\label{coupling1}\right)
\end{eqnarray}
Introduce, the neutral vector bosons $V_{\mu}^{\prime}$ and
$V_{\mu}^{\prime\prime}$
\begin{eqnarray}
\frac{V_{\mu }^{\prime }}{g^{\prime }} &=&\frac{1}{g^{\prime \prime }}%
W_{3R\mu }-\frac{1}{g}\left( \frac{g^{\prime \prime }}{g_{1}}B_{1\mu }+\frac{%
g^{\prime \prime }}{g_{X}}X_{\mu }\right)\notag  \\
\frac{V_{\mu }^{\prime \prime }}{g^{\prime \prime }} &=&\left( \frac{1}{g_{X}%
}B_{1\mu }-\frac{1}{g_{1}}X_{\mu }\label{coup2}\right)
\end{eqnarray}
From Eqs.(\ref{coupling1}) and (\ref{coup2})
\begin{eqnarray}
\frac{1}{e^{2}} &=&\frac{1}{g^{2}}+\frac{1}{g^{\prime 2}}\notag \\
\frac{1}{g^{\prime 2}} &=&\frac{1}{g^{2}}+\frac{1}{g_{1}^{2}}+\frac{1}{%
g_{X}^{2}}\notag \\
&=&\frac{1}{g^{2}}+\frac{1}{g^{\prime \prime 2}}\notag \\
\frac{1}{g^{\prime \prime 2}}
&=&\frac{1}{g_{1}^{2}}+\frac{1}{g_{X}^{2}}
\end{eqnarray}
From the above equations
\begin{eqnarray}
A_{\mu } &=&\frac{e}{g}W_{3L\mu }+\frac{e}{g^{\prime }}B_{\mu }\notag \\
Z_{\mu } &=&\frac{e}{g^{\prime }}W_{3L\mu }-\frac{e}{g}B_{\mu } \\
\frac{e}{g} &=&\sin \theta _{W},\text{ \ }\frac{e}{g^{\prime }}=\cos
\theta _{W},\text{\ }\tan \theta _{W}=\frac{g^{\prime }}{g}
\end{eqnarray}
\begin{eqnarray}
W_{33\mu } &=&\frac{g^{\prime \prime }}{g}\left[ \frac{g}{g^{\prime
}}V_{\mu }^{\prime
}-\frac{1}{3}\frac{2g_{1}^{2}-g_{X}^{2}}{g_{1}g_{X}}V_{\mu }^{\prime
\prime }\right] =Z_{\mu }^{\prime } \notag\\
W_{44\mu }^{\prime \prime } &=&\frac{g^{\prime \prime }}{g}\left[ -\frac{g}{%
g^{\prime }}V_{\mu }^{\prime }-\frac{1}{3}\frac{2g_{1}^{2}+5g_{X}^{2}}{%
g_{1}g_{X}}V_{\mu }^{\prime \prime }\right] =Z_{\mu }^{\prime \prime
}
\end{eqnarray}
The diagonal matrix elements given in Eq.(12) can be expressed in
terms of four physical neutral vector bosons $A_{\mu}$, $Z_{\mu}$,
$Z^{\prime}_{\mu}$ and $Z^{\prime\prime}_{\mu}$ of the group
$SU(4)\times U_{X}(1)$.

In order to give masses to the vector bosons, the gauge symmetry
group $SU(4)\times U_{X}(1)$ is spontaneously broken.The gauge symmetry is spontaneously broken in the following pattern
\begin{eqnarray}
SU(4)\times U_X (1)\xrightarrow[V^2]{}SU_L(2)\times SU_R(2)\times
U_{Y_1(1)}\times U_{X}(1)\notag \\
\xrightarrow{v_3^2,v_4^2}SU_L(2)\times
U_Y(1)\xrightarrow{v^2}U_{em}(1)\notag
\end{eqnarray}
$V^2>>v_3^2,v_4^2>>v^2$.

For this pattern, the first stage of symmetry breaking is
accomplished by 15-plet of Higgs $\eta$:
\begin{eqnarray}
\langle\eta\rangle=\frac{V}{\sqrt{2}}diag(1,1,-1.-1)\notag
\end{eqnarray}
This gives the lepto-quarks mass term:
\begin{eqnarray}
M(lepto-quark)=\frac{1}{2}g^2V^2[2X_1^{2/3\mu}X_{1\mu}^{-2/3}+2X_{2}^{1/3\mu}X_{2\mu}^{-1/3}+2Y_1^{5/3\mu}Y_{1\mu}^{-5/3}+2Y_2^{2/3\mu}Y_{2\mu}^{-2/3}]
\end{eqnarray}
The gauge group $SU_{L}(2)\times SU_{R}(2)\times U_{Y_{1}}(1)\times
U_{X}(1)$ is broken to the standard model group $SU_{L}(2)\times
U_{Y}(1)$ by introducing scalar multiplets $\Phi_{3}$ and
$\Phi_{4}$:

\begin{eqnarray}
\Phi_{3}:(1,2)_{Y_{1}=-2/3, Y_{X}=-1/3} =\left(\begin{array}{c} 0\\
0\\
\Phi_{3}^{0}\\
\Phi_{3}^{-}%
\end{array}%
\right)\to\left(\begin{array}{c} 0\\
0\\
\frac{1}{\sqrt{2}}(v_{3}+H_{3})\\
0%
\end{array}%
\right)
\end{eqnarray}
\begin{eqnarray}
\Phi _{4}:(1,2)_{Y_{1}=-2/3,Y_{X}=5/3}=\left(
\begin{array}{c}
0 \\
0 \\
\Phi _{4}^{+} \\
\Phi _{4}^{0}%
\end{array}%
\right) \rightarrow \left(
\begin{array}{c}
0 \\
0 \\
H_{4}^{+} \\
\frac{1}{\sqrt{2}}\left( v_{4}+H_{4}\right)
\end{array}%
\right)
\end{eqnarray}
The group $SU_{L}(2)\times U_{Y}(1)$ is broken to $U_{em}(1)$ by a
scalar multiplet
\begin{eqnarray}
\Phi :(2,1)_{Y_{1}= 2/3,Y_{X}=1/3}=\left(
\begin{array}{c}
\Phi^{+} \\
\Phi^{0} \\
0 \\
0%
\end{array}%
\right) \rightarrow \left(
\begin{array}{c}
0 \\
\frac{1}{\sqrt{2}}\left( v+H\right) \\
0 \\
0%
\end{array}%
\right)
\end{eqnarray}

Thus the vector boson mass term is given by
\begin{eqnarray}
M_{W} &=&\frac{1}{2}g^{2}v^{2}\left[ 2W_{L}^{+\mu }W_{L\mu
}^{-}+\frac{1}{\cos ^{2}\theta _{W}}Z^{\mu }Z_{\mu }+2X_{2}^{1/3\mu
}X_{2\mu
}^{-1/3}+2Y_{2}^{2/3 \mu }Y_{2 \mu }^{-2/3}\right]  \notag\\
&&+\frac{1}{2}g^{2}v_{3}^{2}\left[ 2X_{1}^{2/3 \mu }X_{1\mu
}^{-2/3}+2X_{2}^{1/3 \mu }X_{2\mu }^{-1/3}+Z^{\mu \prime} Z'_{\mu
}+2W_{R}^{-\mu }W_{R \mu }^{+}\right]  \notag\\
&&+\frac{1}{2}g^{2}v_{4}^{2}\left[ 2Y_{1}^{5/3\mu }Y_{1\mu
}^{-5/3}+2Y_{2}^{2/3 \mu }Y_{2 \mu }^{-2/3}+2W_{R}^{+\mu }W_{R\mu
}^{-}+Z''^{\mu }Z''_{\mu }\right]
\end{eqnarray}

where we have neglected the contribution of $\langle \Phi \rangle $
to $Z^{\prime}_{\mu}$ and $Z^{\prime\prime}_{\mu}$ in the first term
of Eq.(23), since $v_{3}^{2}$, $v_{4}^{2}$ $\gg$ $v^{2}$. For the
multiplets given in Eq.(9), the quarks and leptons acquire mass by
the vacuum expectation value of the scalar multiplets $\Phi$ and
$\Phi_{3}$.

For this pattern of symmetry breaking we get following mass pattern
for the vector bosons
\begin{eqnarray}
m(\text{lepto-quarks})>>m_{W_{R\mu}}^{\pm},Z^{\prime}_{\mu},Z^{\prime\prime}_{\mu}>>m_{W_{L\mu}}^{\pm},Z\notag
\end{eqnarray}

To have an intermediate mass scale
for $W_{R\mu}^{\pm},Z^{\prime}_{\mu},Z^{\prime\prime}_{\mu}$ is attractive, as for this case, the phenomenology is similar to those gauge groups with an extra neutral vector boson $Z^{\prime}$.
\section{Gauge Invariant
Lagrangian}

 For the quark leptons assignment to the representations
$\textbf({4_L},{1_R}), \textbf({4_R},{1_L})$ given in Eq.(9) the
gauge invariant Lagrangian :
\begin{eqnarray}
\mathcal{L} &\mathcal{=}&\bar{F}_{L,R}i\gamma ^{\mu }\left( \partial _{\mu }+%
\frac{i}{2}g\vec{\lambda}\cdot \vec{W}_{\mu }\mp
\frac{i}{6}g_{X}X_{\mu }\right) F_{L,R} \notag\\
&&+\bar{f}_{R}i\gamma ^{\mu }\left( \partial _{\mu }+i\frac{Y_{X}}{2}%
g_{X}X_{\mu }\right) f_{R}+\bar{f}_{L}i\gamma ^{\mu }\left( \partial
_{\mu }+i\frac{Y_{X}^{C}}{2}g_{X}X_{\mu }\right) f_{L}\label{GIL1}
\end{eqnarray}
From Eq.(\ref{GIL1}), the interaction Lagrangian
\begin{eqnarray*}
\mathcal{L}_{\text{int}}=\mathcal{L}_{int}(\text{charged})+\mathcal{L}_{\text{int}}(\text{B-L)}+\mathcal{L}_{int}\text{(neutral})
\end{eqnarray*}
\begin{eqnarray}
\mathcal{L}_{\text{int}}(\text{charged)}=-\frac{g}{2\sqrt{2}}\left\{
\begin{array}{c}
\left[ (\bar{\nu}_{e}\gamma ^{\mu }e+\bar{u}\gamma ^{\mu
}d)_{L}W_{L\mu
}^{-}+h.c.\right]  \\
\left[ +(\bar{\nu}_{E}\gamma ^{\mu }E+\bar{U}\gamma ^{\mu
}D)_{L}W_{R\mu }^{-}+h.c.\right]
\end{array}%
\right\}
\end{eqnarray}
\begin{eqnarray}
\mathcal{L}_{\text{int}}(\text{B-L)}=-\frac{g}{2\sqrt{2}}\left\{
\begin{array}{c}
\left[ (\bar{u}\gamma ^{\mu }E-\bar{U}\gamma ^{\mu }e)_{L}Y_{1\mu
}^{-5/3}+h.c.\right]  \\
+\left[ (\bar{d}\gamma ^{\mu }E+\bar{U}\gamma ^{\mu }\nu
_{e})_{L}Y_{2\mu
}^{-2/3}+h.c.\right]  \\
+\left[ (\bar{u}\gamma ^{\mu }\nu _{E}+\bar{D}\gamma ^{\mu
}e)_{L}X_{1\mu
}^{-2/3}+h.c.\right]  \\
+\left[ (\bar{d}\gamma ^{\mu }\nu _{E}-\bar{D}\gamma ^{\mu }\nu
_{e})_{L}X_{2\mu }^{1/3}+h.c.\right]
\end{array}%
\right\}
\end{eqnarray}

\begin{eqnarray}
 \mathcal{L}_{\text{int}}(\text{neutral})&=&-g\bigg[\sin\theta_W J^\mu
_{em}A_{\mu}+\frac{1}{2\cos\theta_W}J^{\mu}_ZZ_{\mu}\notag \\
&&+\frac1 4 [\tan^2\theta_W(-2J^{\mu}_{em}+J^{\mu}_{3\text{ }V-A})(1-\frac3 2\frac{g^2}{g_1^2})\notag \\
&&+((\nu_e\gamma^{\mu}\nu_e+\overline{e}\gamma^{\mu}e)-(\overline{u}\gamma^{\mu}u+\overline{d}\gamma^{\mu}d)\notag \\
&&+2(\overline{\nu}_E\gamma^{\mu}\nu_E-2\overline{D}\gamma^{\mu}D))_L]Z^{\prime}_{\mu}\notag \\
&&+\frac1 4[\tan^2\theta_W(2J^{\mu}_{em}-J^{\mu}_{3\text{ }V-A})(1+\frac3 2\frac{g^2}{g_1^2})\notag \\
&&+((\nu_e\gamma^{\mu}\nu_e+\overline{e}\gamma^{\mu}e)-(\overline{u}\gamma^{\mu}u+\overline{d}\gamma^{\mu}d)\notag \\
&&+4(\overline{E}\gamma^{\mu}E-\overline{U}\gamma^{\mu}U))_L]Z^{\prime\prime}_{\mu}\bigg]
\end{eqnarray}
\begin{eqnarray}
 J^{\mu}_{em}&=&[-\overline{e}\gamma^{\mu}e+\frac2 3\overline{u}\gamma^{\mu}u-\frac1 3\overline{d}\gamma^{\mu}d\notag \\
&&-\overline{E}\gamma^{\mu}E+\frac2 3\overline{U}\gamma^{\mu}U-\frac1 3\overline{D}\gamma^{\mu}D]\notag \\
J^{\mu}_Z
&=&[J^{\mu}_{3\text{ }V-A}-2\sin^2\theta_WJ^{\mu}_{em}]\notag \\
J^{\mu}_{3\text{ }V-A}&=&[(\overline{\nu}\gamma^{\mu}\nu_e]-\overline{e}\gamma^{\mu}e+(\overline{u}\gamma^{\mu}u-\overline{d}\gamma^{\mu}d))]_L\end{eqnarray}

From Eqs.(25) and (27), it is clear that neither the charged (V-A)
current nor the neutral current for the mirror fermions is coupled
to the vector bosons $W^{\pm}_{L\mu}$ and $Z_{\mu}$. Thus the mirror
fermions are decoupled from the fermions of the standard model
except through lepto quark bosons or through electromagnetic
current. In particular the decays:
\begin{eqnarray*}
\nu _{E}\rightarrow d+\bar{D}+\nu _{e},\text{ \ \ \ }\nu _{E}\rightarrow u+%
\bar{D}+e^{-}
\end{eqnarray*}
through lepto-quarks $X^{1/3}_{2\mu}$ and $X^{-2/3}_{1\mu}$ are not
energetically allowed. Also the decay of $\nu_{E}$ through neutral
vector bosons $Z^{\prime}_{\mu}$, $Z^{\prime\prime}_{\mu}$ is not
possible. The heavy neutrinos $\nu_{E_{i}}$, $i =1,2,3$ are stable
and are good candidates for the dark matter.

Finally the couplings of scalar multiplets to fermions give mass
term :
\begin{eqnarray}
&&\left[ \left( h_{d}\bar{d}_{L}d_{R}+h_{u}\bar{u}_{L}u_{R}+h_{e}\bar{e}%
_{L}e_{R}\right) +h.c\right] \frac{1}{\sqrt{2}}(v+H) \notag\\
&&+\left[ \left( h_{\nu _{E}}\bar{\nu}_{EL}N_{ER}+h_{E}\bar{E}_{L}E_{R}+h_{D}%
\bar{D}_{L}D_{R}+h_{u}\bar{u}_{L}u_{R}\right) +h.c\right] \frac{1}{\sqrt{2}}%
(v_{3}+H_{3}) \notag\\
 &&+\left[ \left(
f_{d}\bar{d}_{L}E_{R}+f_{e}\bar{U}_{R}\nu _{eL}\right)
H^{-2/3}+h.c.\right]
\end{eqnarray}
where an extra right-handed neutrino $N_{ER}$ is introduced. The
neutrino $\nu_{E}$ acquires Majorana mass through see-saw mechanism.
From Eq. (27), it is clear that the decay channels $Z\rightarrow
E^-E^+,U\overline{U},D\overline{D}$ are available to Z-boson through $J_{em}^{\mu}$ in $J_Z^{\mu}$. The decay width of Z boson is
accurately accounted for in terms of the decay channels of the
standard model, it follows that $m_E\geq m_{Z}/2$; the mirror
electron being the lightest fermion coupled to Z. Hence the mass
scale $v_3/v$ for breaking the group $SU_L(2)\times SU_R(2)\times U_{Y_1}(1)\times U_X(1)$ to
$SU_L(2)\times U_Y(1)$ is given by
\begin{eqnarray}
\frac{h_Ev_3}{h_ev}&\geq&\frac{m_Z/2}{m_e}\notag \\
\frac{v_3}{v}&\geq&\frac{m_Z}{2m_e}(\frac{h_e}{h_E})\sim10^5(\frac{h_e}{h_E}).
\end{eqnarray}
Hence from the above equation
\begin{eqnarray}
m^\prime_Z/m_Z\geq10^5(\frac{h_e}{h_E})\notag
\end{eqnarray}

If $h_e\sim h_E$, $m_Z^{\prime}\geq10^5m_Z$. However, if $h_E>>
h_e$, then the lower limit for $m_{Z^{\prime}}$ is much lower. For
example, $h_E/h_e~10^{3}$, $m_{Z^{\prime}}\geq10^2m_Z$. The mass hierarchy of fermions in the standard model requires that Yukawa coupling of Higgs scalar increases with mass of the fermions; hence assuming this pattern, it is possible $h_E>>h_e$. If $v_4>v_3$, then $Z^{\prime}$ is lighter than $Z^{\prime\prime}$; it has similar phenomenology as for the models with an extra neutral vector boson. If $v_4\approx v_3$, then we have two extra neutral vector bosons degenerate in mass.

\section{Conclusion and summary}
We briefly discuss the baryon genesis. A typical lepto-quark decays
into a mirror fermion accompanied by an ordinary light antifermion:
\begin{eqnarray*}
\bar{X} &\rightarrow &\bar{q}L:r_{\bar{q}}\text{ \ \ \ \ \ \ \ \ \ \ \ \ \ }%
X\rightarrow q\bar{L}:r_{q} \\
&\rightarrow &\bar{Q}l:r_{\bar{Q}}\text{ \ \ \ \ \ \ \ \ \ \ \ \ \ \ }%
\rightarrow Q\bar{l}:\,r_{Q}
\end{eqnarray*}
where
\begin{eqnarray*}
r_{\bar{q}}+r_{\bar{Q}}=1,\text{ \ \ }r_{q}+r_{Q}=1
\end{eqnarray*}

As is well known, baryon genesis requires

\begin{eqnarray*}
r_{q}\neq r_{\bar{q}},\text{ \ }r_{Q}\neq r_{\bar{Q}}
\end{eqnarray*}

For baryon genesis in our universe
\begin{eqnarray*}
\Delta B_{q} &=&\left( B_{q}-B_{\bar{q}}\right) =\left(
r_{q}-r_{\bar{q}}\right) \qquad r_{q}>r_{\bar{q}} \\
\Delta L&=&(L-\overline{L})<0
\end{eqnarray*}

Now
\begin{eqnarray*}
\Delta B_{Q} &=&\left( B_{Q}-B_{\bar{Q}}\right) =\left( r_{Q}-r_{\bar{Q}%
}\right)  \\
&=&\left[ \left( 1-r_{q}\right) -\left( 1-r_{\bar{q}}\right) \right]
=\left( r_{\bar{q}}-r_{q}\right)<0 \\
\Delta l&=&(l-\overline{l})>0
\end{eqnarray*}
Hence for our Universe; $\Delta B_{q}>0$, $\Delta l>0$, implies
$\Delta B_{Q} < 0$; $\Delta L< 0$

Hence baryon-genesis in our universe implies antibaryon genesis in
the mirror universe; overall a symmetric Universe.

To summarize: A model for electroweak unification of quarks and
leptons is constructed. The gauge group for unification is
$SU_C(3)\times SU(4)\times U_X(1)$, which is just an extension of
$SU_L(2)\times U(1)$ to $SU(4)\times U_X(1)$. The model requires
three generations of quarks and leptons which are replica (mirror)
of quarks and leptons of the standard model. The gauge group
$SU(4)\times U_X(1)$ is spontaneously broken in such a way as to
reproduce the standard model and to generate mass pattern
$m(\text{lepto-quarks})>>m^{\pm}_{W_{R\mu}},Z^{\prime}_{\mu},Z^{\prime\prime}_{\mu}>>m^{\pm}_{W_{L\mu}},Z_{\mu}$.
The Z-boson decay width give the lower limit, on the mass scale of
mirror fermions: $m_E\geq m_Z/2$, the lightest mirror fermion
coupled to Z. The lepto-quarks connect the
standard model and mirror fermions. The matter dominated our
Universe implies antimatter dominated mirror Universe.

\end{document}